\documentclass[acus]{JAC2003}

\usepackage{graphicx}
\usepackage{epsfig}
\usepackage{axodraw}

\newcommand{\lsim}
{\mathrel{\raisebox{-.3em}{$\stackrel{\displaystyle <}{\sim}$}}}
\newcommand{\gsim}
{\mathrel{\raisebox{-.3em}{$\stackrel{\displaystyle >}{\sim}$}}}
\def\asymp#1%
{\mathrel{\raisebox{-.4em}{$\widetilde{\scriptstyle #1}$}}}

\def\Nequal#1%
{\mathrel{\raisebox{-.5em}{$\stackrel{=}{\scriptstyle\rm#1}$}}}
\newcommand{\dsl}[1]{\not \hspace{-0.7mm}#1}
\def\dsl{\mathpalette\make@slash}
\def\make@slash#1#2{\setbox\z@\hbox{$#1#2$}%
  \hbox to 0pt{\hss$#1/$\hss\kern-\wd0}\box0}

\def\beq{\begin{equation}}
\def\eeq{\end{equation}}
\def\beqar{\begin{eqnarray}}
\def\eeqar{\end{eqnarray}}
\def\barr#1{\begin{array}{#1}}
\def\earr{\end{array}}
\def\bfi{\begin{figure}}
\def\efi{\end{figure}}
\def\btab{\begin{table}}
\def\etab{\end{table}}
\def\bce{\begin{center}}
\def\ece{\end{center}}
\def\nn{\nonumber}
\def\disp{\displaystyle}
\def\text{\textstyle}

\def\ga{\gamma}
\def\de{\delta}

\def\eps{\epsilon}

\def\la{\lambda}

\def\si{\sigma}

\def\reffi#1{\mbox{Figure~\ref{#1}}}

\def\refta#1{\mbox{Table~\ref{#1}}}

\def\citere#1{\mbox{Ref.~\cite{#1}}}
\def\citeres#1{\mbox{Refs.~\cite{#1}}}

\newcommand{\TeV}{\unskip\,\mathrm{TeV}}
\newcommand{\GeV}{\unskip\,\mathrm{GeV}}
\newcommand{\MeV}{\unskip\,\mathrm{MeV}}
\newcommand{\pba}{\unskip\,\mathrm{pb}}
\newcommand{\fb}{\unskip\,\mathrm{fb}}

\newcommand{\ri}{{\mathrm{i}}}
\newcommand{\rd}{{\mathrm{d}}}

\newcommand{\Oa}{\mathswitch{{\cal{O}}(\alpha)}}

\newcommand{\M}{{\cal{M}}}
\def\mathswitchr#1{\relax\ifmmode{\mathrm{#1}}\else$\mathrm{#1}$\fi}
\newcommand{\Pf}{\mathswitch  f}

\newcommand{\PW}{\mathswitchr W}
\newcommand{\Pw}{\mathswitchr w}
\newcommand{\PZ}{\mathswitchr Z}
\newcommand{\PA}{\mathswitchr A}
\newcommand{\Pg}{\mathswitchr g}
\newcommand{\Ph}{\mathswitchr h}
\newcommand{\PH}{\mathswitchr H}
\newcommand{\Pe}{\mathswitchr e}

\newcommand{\Pne}{\mathswitch \nu_{\mathrm{e}}}
\newcommand{\Pnebar}{\mathswitch \bar\nu_{\mathrm{e}}}
\newcommand{\Pd}{\mathswitchr d}

\newcommand{\Pu}{\mathswitchr u}

\newcommand{\Pb}{\mathswitchr b}

\newcommand{\Pt}{\mathswitchr t}

\newcommand{\Pep}{\mathswitchr {e^+}}
\newcommand{\Pem}{\mathswitchr {e^-}}

\def\mathswitch#1{\relax\ifmmode#1\else$#1$\fi}

\newcommand{\MW}{\mathswitch {M_\PW}}

\newcommand{\MZ}{\mathswitch {M_\PZ}}
\newcommand{\MH}{\mathswitch {M_\PH}}
\newcommand{\Me}{\mathswitch {m_\Pe}}

\newcommand{\Mt}{\mathswitch {m_\Pt}}

\newcommand{\sw}{\mathswitch {s_\Pw}}
\newcommand{\cw}{\mathswitch {c_\Pw}}

\newcommand{\GF}{\mathswitch {G_\mu}}

\def\solid{\raise.9mm\hbox{\protect\rule{1.1cm}{.2mm}}}
\def\dash{\raise.9mm\hbox{\protect\rule{2mm}{.2mm}}\hspace*{1mm}}

\newcommand{\eennh}{\Pep\Pem\to\nu\bar\nu\PH}
\newcommand{\eetth}{\Pep\Pem\to\Pt\bar\Pt\PH}

\begin{document}

\onecolumn
\setcounter{page}{0}
{\Large
\strut\hfill MPP-2003-46 \\
\strut\hfill hep-ph/0308079 \\[3cm]
\begin{center}
{\Large\bf\boldmath{THEORETICAL TOOLS \\[.5em]
FOR A FUTURE $\Pep\Pem$ LINEAR COLLIDER}}%
\footnote{To appear in the proceedings of {\it The 4th ECFA/DESY
Workshop on Physics and Detectors for a 90-800 GeV Linear $e^+e^-$ Collider},
NIKHEF, Amsterdam, the Netherlands, April 2003.}
\\[1cm]
Stefan Dittmaier \\[1cm]
{\it Max-Planck-Institut f\"ur Physik (Werner-Heisenberg-Institut) \\[.5em]
F\"ohringer Ring 6, D-80805 M\"unchen, Germany}
\end{center}\par
\vskip 4cm {\bf Abstract:} \par
Recent progress in the calculation of radiative corrections
and in Monte Carlo event generation, relevant for a future $\Pep\Pem$ 
linear collider, is reviewed.
\par \vfill \noindent July 2003 \\[1em] \null }
\twocolumn
\clearpage

\title{\boldmath{THEORETICAL TOOLS FOR A FUTURE $\Pep\Pem$ LINEAR COLLIDER%
\thanks{%
The work reported in this article is based on the progress reached since
Spring 2001 by the ``Loopverein'' and ``Monte Carlo event generators'' 
working groups of the extended ECFA/DESY Study; 
particular thanks goes to the working-group members
M.~Antonelli (Frascati),
M.~Awramik (Cracow), 
D.~Bardin (Dubna),
G.~B\'elanger (Annecy),
A.~Biernacik (Katowice),
J.~Bl\"umlein (DESY Zeuthen),
F.~Boudjema (Annecy),
A.~Brandenburg (Aachen),
C.~M.~Carloni Calame (Pavia),
P.~Ciafaloni (Lecce),
M.~Czakon (Katowice),
A.~Denner (PSI Villigen),
M.~Diaz (Catolica),
M.~Faisst (Karlsruhe),
J.~Fleischer (Bielefeld), 
J.~Fujimoto (KEK),
A.~Freitas (FNAL), 
F.~Gangemi (Pavia),
A.~Ghinculov (Rochester), 
P.~Golonka (Cracow),           
J.~Guasch (PSI Villigen),
T.~Hahn (MPI Munich), 
A.~van Hameren (Athens),
S.~Heinemeyer (LMU Munich),
W.~Hollik (MPI Munich),
V.~A.~Ilyin (Moscow),  
T.~Ishikawa (KEK),
S.~Jadach (Cracow), 
F.~Jegerlehner (DESY Zeuthen),
L.~Kalinovskaya (Dubna), 
M.~Kalmykov (DESY Zeuthen),
T.~Kaneko (KEK),
K.~Kato (Tokyo),
H.~Kawamura (DESY Zeuthen),
W.~Kilian (DESY Hamburg),
K.~Ko\l{}odziej (Katowice),
M.~Kr\"amer (Edinburgh), 
F.~A.~Krauss (CERN),
J.~K\"uhn (Karlsruhe), 
Y.~Kurihara (KEK),
M.~Maniatis (Hamburg),
K.~M\"onig (DESY Zeuthen),
G.~Montagna (Pavia),
M.~Moretti (Ferrara),
S.~Moretti (Southampton),
O.~Nicrosini (Pavia),
A.~Leike (LMU Munich), 
A.~Lorca (Granada), 
T.~Ohl (W\"urzburg),     
W.~\"Oller (Vienna),
A.~I.~Onishchenko (Moscow),
C.~Papadopoulos (Athens),
G.~Passarino (Torino),
M.~Peskin (SLAC),
F.~Piccinini (CERN/Pavia),
T.~Pierzchala (Karlsruhe),
W.~P\l{}aczek (Cracow),
T.~Riemann (DESY Zeuthen), 
M.~Ronan (Berkeley),
M.~Roth (Karlsruhe),
C.~Schappacher (Karlsruhe), 
S.~Schumann (Dresden),
Y.~Shimizu (KEK),
M.~Skrzypek (Cracow),
M.~Spira (PSI Villigen),
O.~Tarasov (DESY Zeuthen),
B.~Tausk (Freiburg), 
O.~Veretin (Karlsruhe),
C.~Verzegnassi (Trieste), 
D.~Wackeroth (Buffalo),
B.~F.~L.~Ward (Knoxville),
Z.~Was (Cracow),
C.~Weber (Vienna), 
M.~Weber (PSI Villigen),
G.~Weiglein (Durham),
S.~Weinzierl (Parma), 
A.~Werthenbach (CERN),
M.~Wing (Bristol),
M.~Worek (Katowice),
P.~Zerwas (DESY Hamburg).
}}}

\author{S.~Dittmaier, Max-Planck-Institut f\"ur Physik
(Werner-Heisenberg-Institut), Munich, Germany%
\thanks{Stefan.Dittmaier@mppmu.mpg.de}}

\maketitle

\begin{abstract}
Recent progress in the calculation of radiative corrections
and in Monte Carlo event generation, relevant for a future $\Pep\Pem$ 
linear collider, is reviewed.
\end{abstract}

\section{INTRODUCTION}

Precision measurements at LEP, SLC, and the Tevatron rendered the last decade
the era of high-precision physics.
A future $\Pep\Pem$ linear collider (LC), such as TESLA 
\cite{Aguilar-Saavedra:2001rg}, the NLC \cite{Abe:2001wn}, or the GLC 
(former JLC)
\cite{Abe:2001gc}, does not only offer an even greater physics potential, 
but in turn represents a 
great challenge for theorists to understand phenomena at 
the experimentally achievable level of precision.

For instance, returning again to the Z-boson resonance in the
``GigaZ'' mode of TESLA (where about $10^9$ Z bosons can be produced
within 50--100 days of running) allows for a repetition
of the LEP1/SLC physics program with roughly an order of magnitude
higher precision
(see also \citere{Baur:2001yp}).
Specifically, the uncertainty in the effective weak 
mixing angle could be reduced from $1.7\times 10^{-4}$ to 
$1.3\times 10^{-5}$.
For a theoretical description of the Z~resonance at this level of accuracy 
full two-loop calculations of the observables as well as the knowledge
of leading higher-order effects are clearly necessary.
A scan over the W-pair production threshold could provide a sensitivity to
the W-boson mass of about $7\MeV$, which should be compared with the
present error of $34\MeV$, resulting from the kinematical reconstruction
of W~bosons. The present approach of approximating
the radiative corrections to $\Pep\Pem\to\PW\PW\to 4\,$fermions
by an expansion about the 
double resonance is not applicable (or at least not reliable) in the
threshold region where singly- or non-resonant contributions become
important. The only solution seems to be the full treatment of the
complete four-fermion production processes at the one-loop level,
including higher-order improvements. 

At energies exceeding the reach of LEP2, many new processes will
be accessible, such as top-quark pair production, Higgs production
(if the Higgs boson exists), or reactions with new-physics
particles, as e.g.\ predicted by SUSY models. Most of these
heavy particles are unstable, so that their production eventually leads
to many-particle final states. For example, the production of
$\Pt\bar\Pt$ pairs or of a Higgs-boson with an intermediate or large mass
($\MH>2\MW$) leads to six-fermion final states. To exploit the 
potential of a LC, predictions for such reactions should be based
on full transition matrix elements and improved by radiative corrections
as much as possible. 
The higher level of accuracy at a future LC does, however, not only call for
proper event generators for many-particle final states. ``True'' event
generators, i.e.\ including parton showering and hadronization, have to
be improved as well.

In this brief article the main progress on precision calculations and 
event generators that has been achieved
in the ``Loopverein'' and ``Generators'' working groups of
the Extended ECFA/DESY Study since the appearance of the TESLA TDR
\cite{Aguilar-Saavedra:2001rg} is reviewed.
More studies on the physics potential of a LC in view of 
electroweak and strong interactions, top-quark physics, Higgs physics,
and new physics searches, in particular supersymmetry,
are summarized in \citere{theseprocs}.

\section{HIGH-PRECISION OBSERVABLES AND MULTI-LOOP CALCULATIONS}

\subsection{Precision calculations for $\mu$ decay}

The precision measurement of the muon lifetime, or equivalently of the
Fermi constant $\GF$, sets an important constraint on the SM parameters,
\beq
\GF = \frac{\pi\alpha(0)}{\sqrt{2}\MW^2\sw^2}\,(1+\Delta r),
\eeq
where $\sw^2=1-\cw^2=1-\MW^2/\MZ^2$ and
the quantity $\Delta r$ comprises the radiative corrections to 
muon decay (apart from the photonic corrections in the Fermi model).
In the past it has become common practice to implicitly solve this relation
for the W-boson mass $\MW$, thus yielding a precise prediction for $\MW$
that can be compared with the directly measured value.
Recently the full prediction at the two-loop level has been completed.
In detail, first the contributions from closed fermion loops and from
bosonic loops involving Higgs-boson exchange were calculated in
\citere{Freitas:2000gg} by making use of the {\sc FeynArts} package
\cite{Kublbeck:1990xc}
and the program {\sc TwoCalc} \cite{Weiglein:hd}
both written in {\sc Mathematica}.
The algebraic reduction leads to two-loop master integrals which are
evaluated by semi-analytical methods. The full bosonic corrections
have been calculated in \citeres{Awramik:2002wn} and 
\cite{Onishchenko:2002ve}; \citere{Awramik:2002wn} includes also
a recalculation of the fermion-loop correction.
In the former calculation the diagrams were generated with the 
{\sc C++} library {\sc Diagen} (by Czakon) and evaluated using 
semi-analytical methods. In the latter case the graphs were
generated with the package {\sc Diana} \cite{Tentyukov:1999is} 
and evaluated by asymptotic expansions. The results of
\citeres{Freitas:2000gg,Awramik:2002wn,Onishchenko:2002ve}
are in good numerical agreement \cite{Awramik:2002vu}.%
\footnote{The results of \citere{Freitas:2000gg} for $\MW$ have
been corrected at the level of $\sim 1\MeV$ recently.}
The two-loop fermionic corrections influence the $\MW$
prediction at the level of $\sim 50\MeV$, where the bulk of this
effect is due to universal, top-mass enhanced corrections to the 
$\rho$-parameter, which are proportional to $\Mt^4$ or $\Mt^2$. 
The non-universal two-loop fermionic corrections have an impact of
up to $4\MeV$, the two-loop bosonic corrections of only $1{-}2\MeV$.

The predictions at the two-loop level have been 
further improved by universal
higher-order corrections to the $\rho$-parameter. The 
corrections of ${\cal O}(\GF^2\Mt^4\alpha_{\mathrm{s}})$ and
${\cal O}(\GF^3\Mt^6)$ have been calculated for arbitrary $\MH$
in \citere{Faisst:2003px}
(for other universal corrections to $\Delta\rho$ and $\Delta r$
see references therein) and were found to change $\MW$ at the level
of $5\MeV$ and $0.5\MeV$, respectively.
The Feynman diagrams were generated using 
{\sc Qgraf} \cite{Nogueira:1991ex} and asymptotically expanded
with the program {\sc Exp} \cite{Harlander:1997zb};
the resulting massive three-loop tadpole integrals were evaluated
with {\sc Matad} \cite{Steinhauser:2000ry}.

\begin{figure}
{ \setlength{\unitlength}{1cm}
\begin{picture}(8,8.0)
\put(-0.5,-0.9){\includegraphics{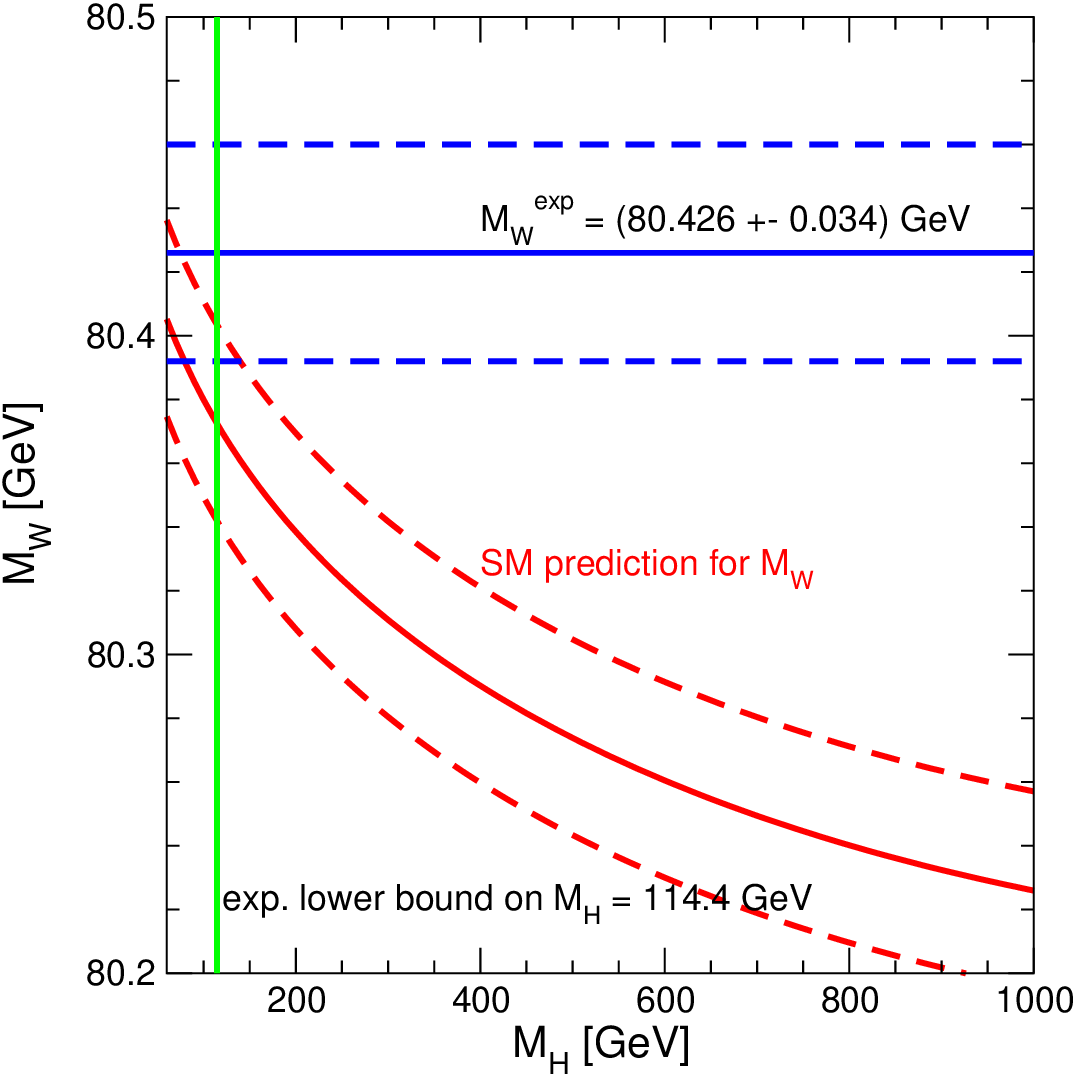}}
\end{picture} }
\caption{Theoretical prediction of the W-boson mass
with the parametric error from uncertainties in the top-quark mass and the 
running electromagnetic coupling, in comparison with the experimental
value $\MW^{\mathrm{exp}}$ (plot shown in \citere{talkGW})}
\label{fig:mwpred}
\end{figure}

Figure~\ref{fig:mwpred} compares the prediction for $\MW$,
including the above-mentioned two-loop and leading three-loop effects,
with the experimental value. 
Note that the shown parametric uncertainty
is much larger than the estimated theoretical uncertainty, which
is about $3{-}4\MeV$ \cite{talkGW,HollikZeuthenprocs}.
Comparing this estimate with the aimed precision of
$7\MeV$ in the $\MW$ determination at a future LC, the prediction of
the W-boson mass from muon decay is in rather good shape.

\subsection{Precision observables on the Z~resonance}

In order to describe the Z-boson resonance at LEP1 within satisfactory
precision it was possible to parametrize the cross section 
near the resonance in such a way \cite{zresonance}
that a Born-like form with generalized
``effective'' couplings is convoluted with QED structure functions
modeling initial-state radiation (ISR).
From these effective Z-boson--fermion couplings so-called 
``pseudo-observables'' were derived, such as various asymmetries, the
hadronic Z-peak cross section, partial Z-decay widths, etc.
Following the formal tree-level parametrization of the couplings, 
an ``effective weak mixing angle'', usually given as
$\sin^2\theta_{f,\mathrm{eff}}$, was derived for each fermion.
Among these parameters the leptonic variable
$\sin^2\theta_{\mathrm{lep,eff}}$ plays a particularly important role,
since it is measured with the high accuracy of $1.7\times 10^{-4}$
and is very sensitive to the Higgs-boson mass.
The state-of-the-art in the precision calculations of the 
pseudo-observables, which is implemented in the programs
{\sc Zfitter} and {\sc Topaz0} (see \citere{Bardin:1998nm}
and references therein), did not change very much since
the release of the TESLA TDR \cite{Aguilar-Saavedra:2001rg}.
For instance, the estimated theoretical uncertainty in
$\sin^2\theta_{\mathrm{lep,eff}}$ is still $\sim 6\times 10^{-5}$.
A critical overview about high-precision physics at the Z~pole, 
in particular focusing on the theoretical uncertainties, can be 
found in \citere{zeuthenMHworkshop}
(see also \citere{Baur:2001yp}).

Whether the pseudo-observable
approach will also be sufficient for Z-boson physics
at the high-luminosity GigaZ option remains to be investigated
carefully. In any case, tremendous theoretical progress will be
needed to match the aimed GigaZ precision on the theoretical side.
For example, the expected experimental accuracy in 
$\sin^2\theta_{\mathrm{lep,eff}}$ is about $1.3 \times 10^{-5}$,
i.e.\ about a factor 4 below the present theoretical uncertainty.
A full control of observables at the two-loop level, improved by
leading higher-order effects, seems to be indispensable.

\subsection{Recent results from the 2-loop frontier}

Although there are no complete next-to-next-to-leading (NNLO)
predictions for 
$2\to2$ scattering reactions and $1\to3$ decays (with one truly
massive leg) available yet, enormous progress was reached in this
direction in recent years. 

Complete virtual two-loop amplitudes for 
(massless) Bhabha scattering \cite{Bern:2000ie},
light-by-light scattering \cite{Bern:2001dg},
and $\Pep\Pem\to3\,$jets \cite{Garland:2001tf}
have been worked out, using a large variety of special techniques,
which have been summarized in \citere{Gehrmann:2002yr}.
A survey of similar results relevant for hadron-collider physics
can also be found there.
Apart from this two-loop progress on massless particle scattering,
also a first step has been made towards massive Bhabha scattering
in \citere{Fleischer:2002wa}.

Full NNLO calculations have to include real double-parton
bremsstrahlung as well as interference contributions of
one-parton bremsstrahlung and one-loop diagrams. The major
complication in these parts concerns the proper extraction of
the infrared (soft and collinear) singularities.
The general form of multiple singular particle emission has been
worked out in \citere{Kosower:2002su}, which can serve
as a basis for the extraction of the singularities.
The actual separation of the singularities can be performed
either by applying phase-space cuts (``slicing approach'')
or by subtracting an auxiliary cross section with the same singular
structure as the original integrand (``subtraction approach'').
In \citere{Weinzierl:2003fx} subtraction terms have been constructed
for the leading colour contribution to $\Pep\Pem\to2\,$jets in NNLO.
However, suitable general subtraction terms as well as their integrated
counterparts that have to be added again are not yet available.

\subsection{Electroweak radiative corrections at high energies}

Electroweak corrections far above the electroweak scale, e.g.\ in the
TeV range, are dominated by soft and collinear gauge-boson exchange,
leading to corrections of the form $\alpha^N\ln^M(s/\MW^2)$ with
$M\le2N$. The leading terms ($M=2N$) are called Sudakov logarithms.
At the one-loop ($N=1$) and two-loop ($N=2$) level 
the leading and subleading corrections to a $2\to2$ process
at $\sqrt{s}\sim 1\TeV$ typically amount to \cite{Denner:2003wi}
\beqar
\de_{\mathrm{LL}}^{\mathrm{1-loop}} &\sim&
-\frac{\alpha}{\pi\sw^2}\ln^2\bigl(\frac{s}{\MW^2}\bigr)  \simeq -26\%, 
\nn\\
\disp \de_{\mathrm{NLL}}^{\mathrm{1-loop}} &\sim&
+\frac{3\alpha}{\pi\sw^2}\ln\bigl(\frac{s}{\MW^2}\bigr) \simeq 16\%,
\nn\\
\disp \de_{\mathrm{LL}}^{\mathrm{2-loop}} &\sim&
+\frac{\alpha^2}{2\pi^2\sw^4}\ln^4\bigl(\frac{s}{\MW^2}\bigr) \simeq 3.5\%,
\nn\\
\disp \de_{\mathrm{NLL}}^{\mathrm{2-loop}} &\sim&
-\frac{3\alpha^2}{\pi^2\sw^4}\ln^3\bigl(\frac{s}{\MW^2}\bigr) \simeq -4.2\%,
\eeqar
revealing that these corrections become significant in the high-energy phase
of a future LC. In contrast to QED and QCD, where the Sudakov logarithms
cancel in the sum of virtual and real corrections, these terms need not
compensate in the electroweak SM for two reasons. 
The weak charges of quarks, leptons, and electroweak gauge bosons
are open, not confined, i.e.\ there is (in contrast to QCD) no need 
to average or to sum over gauge multiplets in the initial or final states
of processes. Even for final states that are inclusive with respect to the
weak charges Sudakov logarithms do not completely cancel owing to the
definite weak charges in the initial state \cite{Ciafaloni:2000gm}.
Moreover, the large 
W- and Z-boson masses make an experimental discrimination of real W- or
Z-boson production possible, in contrast to unobservable soft-photon or
gluon emission.

In recent years several calculations of these high-energy logarithms
have been carried out in the Sudakov regime, where all kinematical
invariants $(p_i p_j)$ of different particle momenta $p_i$, $p_j$
are much larger than all particle masses.%
\footnote{Note that this regime does not cover the case of 
forward scattering of particles, which is also of interest in several cases.}
A complete analysis of all leading and subleading logarithms at the
one-loop level can be found in \citere{Denner:2000jv}.
Diagrammatic calculations of the leading two-loop Sudakov logarithms
have been carried out in \citeres{Denner:2003wi,Beenakker:2000kb}.
Diagrammatic results on the so-called ``angular-dependent''
subleading logarithms have been presented in \citere{Denner:2003wi}.
All these explicit results are compatible with proposed resummations
\cite{Fadin:1999bq,Kuhn:2001hz}
that are based on a symmetric SU(2)$\times$U(1) theory at high energies
matched with QED at the electroweak scale. In this ansatz,
improved matrix elements $\M$ result from lowest-order matrix elements
$\M_0$ upon dressing them with (operator-valued) exponentials,
\beq
\M \sim \M_0 \otimes \exp\left({\de_{\mathrm{ew}}}\right) \otimes
\exp\left({\de_{\mathrm{em}}}\right).
\eeq
Explicit expressions for the electroweak and electromagnetic
corrections $\de_{\mathrm{ew}}$ and $\de_{\mathrm{em}}$, which
do not commute with each other, can, for instance, be found
in \citere{Denner:2003wi}.
For $2\to2$ neutral-current processes of four massless fermions,
even subsubleading logarithmic corrections have been derived and
resummed \cite{Kuhn:2001hz}
using an infrared evolution equation that follows the pattern of QCD.

In supersymmetric models the form of radiative corrections at high energies
has also been worked out for a broad class of processes 
\cite{Beccaria:2001an}. Based on 
one-loop results their exponentiation has been proposed.

\subsection{Higher-order initial-state radiation}

Photon radiation off initial-state electrons and positrons leads to
large radiative corrections of the form $\alpha^N\ln^N(\Me^2/s)$.
These logarithmic corrections are universal and governed by the
DGLAP evolution equations. The solution of these equations for the
electron-photon system yields so-called structure functions, 
generically denoted by $\Gamma(x)$ below,
which can be used via convolution to improve hard scattering cross sections 
$\hat\sigma(p_{\Pe^+},p_{\Pe^-})$ by photon emission effects,
\beqar
\sigma(p_{\Pe^+},p_{\Pe^-}) &=& \int_0^1 \rd x_+\,\Gamma(x_+)
\int_0^1 \rd x_-\,\Gamma(x_-)
\nn\\*
&& {} \times \hat\sigma(x_+p_{\Pe^+},x_-p_{\Pe^-}).
\eeqar
While the soft-photon part of the structure functions 
\mbox{($x\to1$)} can be 
resummed, resulting in an exponential form, the contributions of
hard photons have to be calculated order by order in perturbation theory.
In \citere{Beenakker:1996kt} the structure functions are summarized
up to ${\cal O}(\alpha^3)$. 
\citere{Blumlein:2002bg} describes a new calculation of the
(non-singlet) contributions up to ${\cal O}(\alpha^5)$ and
of the small-$x$ terms $[\alpha\ln^2(x)]^N$ to all orders
(for previous calculations see papers cited in \citere{Blumlein:2002bg}).

\section{\boldmath{RADIATIVE CORRECTIONS TO $2\to3,4,\dots$ PROCESSES}}

\subsection{W-pair production and four-fermion final states}

The theoretical treatment and the presently gained level in accuracy in the 
description of W-pair-mediated $4f$ production were triggered by
LEP2, as it is reviewed in \citeres{Beenakker:1996kt,Grunewald:2000ju}. 
The $\PW$ bosons are treated as resonances in the full
$4f$ processes, $\Pe^+ \Pe^- \to 4 \Pf\, (+\,\gamma)$.
Radiative corrections are split into universal and non-universal corrections.  
The former comprise leading-logarithmic corrections from
ISR, higher-order corrections included by
using appropriate effective couplings, and the Coulomb singularity.  
These corrections can be combined with the full lowest-order matrix elements
easily.
The remaining corrections are called non-universal, since they depend on
the process under investigation.
For LEP2 accuracy, it was sufficient to include these corrections
by the leading term of an expansion about the two $\PW$ poles, defining
the so-called double-pole approximation (DPA). 
Different versions of such a DPA
have been used in the literature
\cite{DPAversions,Jadach:1998tz,Denner:2000kn}.  Although several
Monte Carlo programs exist that include universal corrections, only
two event generators, {\sc YFSWW} \cite{Jadach:1998tz}
and {\sc RacoonWW} \cite{Denner:2000kn,Denner:1999gp},
include non-universal corrections.  

In the DPA approach, the W-pair cross section can be predicted within
$\sim0.5\% (0.7\%)$ in the energy range between $180\GeV$ ($170\GeV$)
and $\sim 500\GeV$, which was sufficient for the LEP2 accuracy of
$\sim 1\%$ for energies $170{-}209\GeV$. At threshold 
($\sqrt{s}\lsim 170\GeV$), the present 
state-of-the-art prediction results from an improved Born approximation
based on leading universal corrections only, because the DPA is not
reliable there, and thus possesses an intrinsic uncertainty of
about $2\%$. In \reffi{fig:mwscan} 
this uncertainty is compared with the sensitivity of the
W-pair production cross section to the W-boson mass and some
assumed experimental data points of a threshold scan. 
\bfi
\centerline{
  \setlength{\unitlength}{.5cm}
  \begin{picture}(14.5,16)
  \put(-0.5,0){\epsfig{file=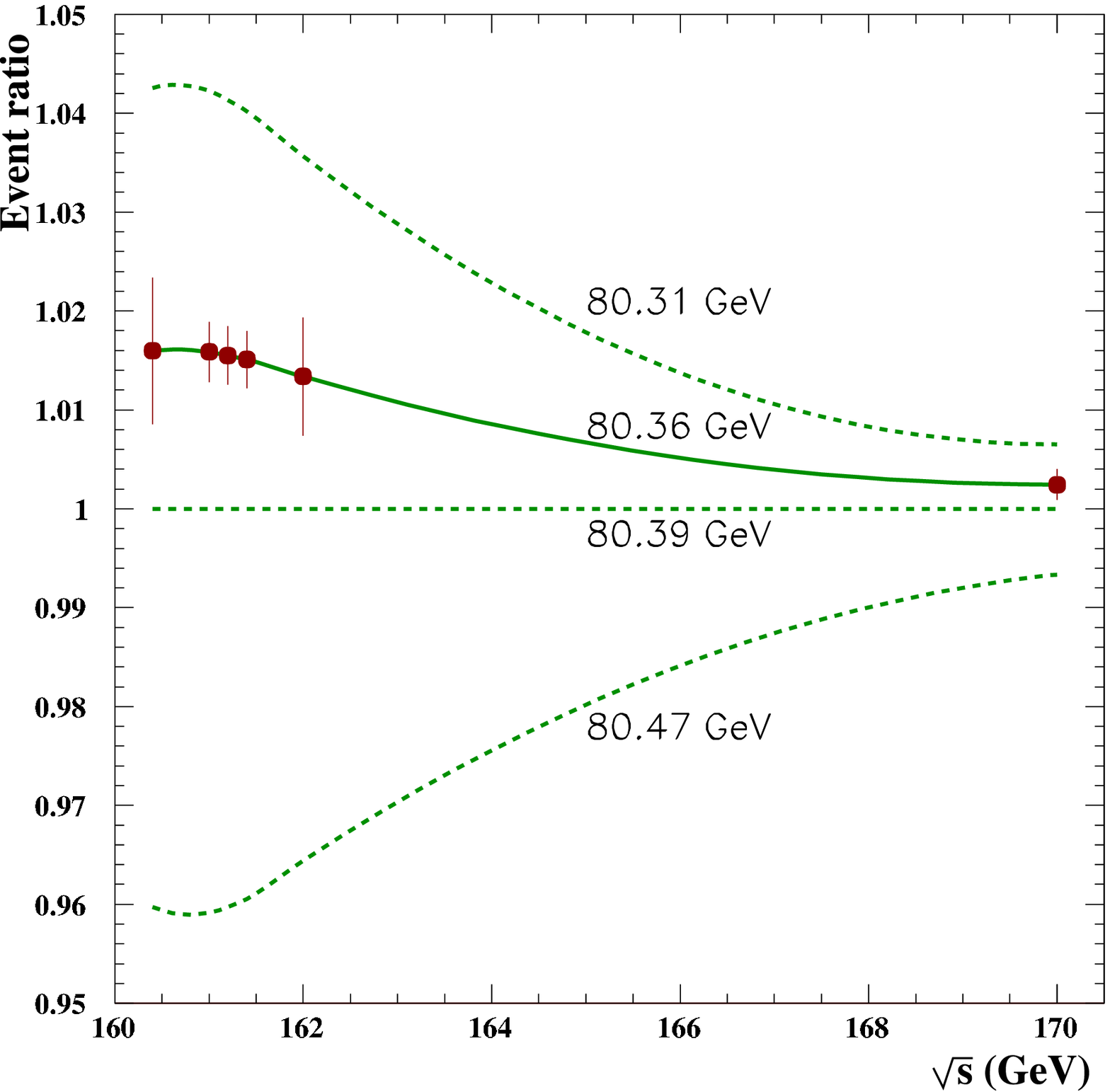, scale=.425}}
  \put( 1.07,10.73){{\line(1,0){13.6}}}
  \put( 1.07,5.3){{\line(1,0){13.6}}}
%  \put(6,13){\tiny{TESLA-TDR '01}}
%  \put(-0.72,-1.0){\Red{\line(1,0){1}}}
%  \put( 1.00,-1.1){$\pm 2\%$ TU on cross section}
  \end{picture} }
\caption{Sensitivity of the
W-pair production cross section to the W-boson mass and some
assumed experimental data points, compared with the theoretical 
uncertainty of $\sim 2\%$ (taken from \citere{Aguilar-Saavedra:2001rg})}
\label{fig:mwscan}
%\efi
\vspace*{4em}
%\bfi
\setlength{\unitlength}{1cm}
\centerline{
\begin{picture}(7,7.6)
\put(-1.8,-15.5){\includegraphics{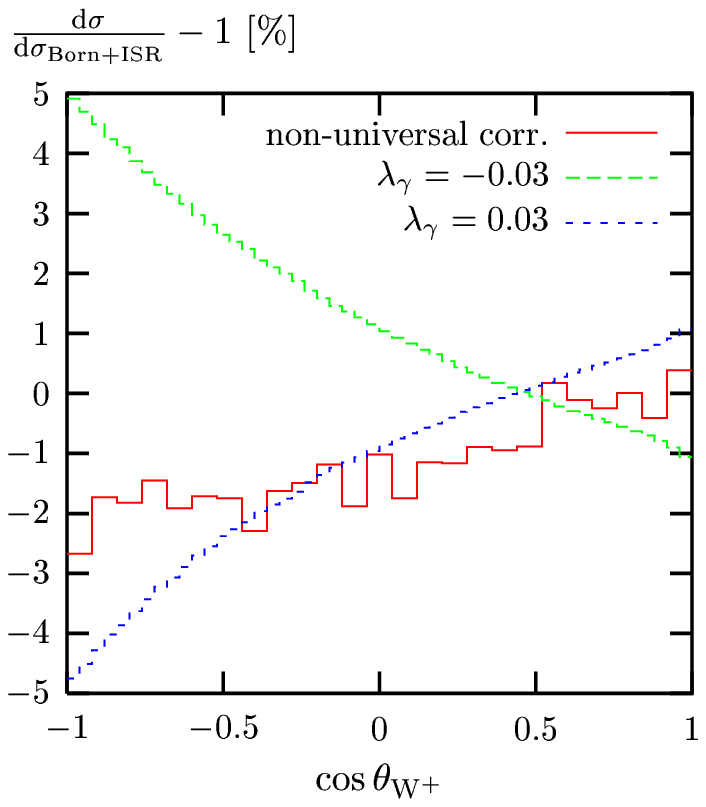}}
\put(4,1.4){{\sc RacoonWW}}
%\put(8,4.4){$\Pep\Pem\to\Pu\bar\Pd\mu^-\bar\nu_\mu$}
%\put(8,3.6){$\sqrt{s}=200\GeV$}
%\put(8,2.8){$\lambda_\PZ=\lambda_\gamma$}
%\put(8,2){$\Delta\kappa_\PZ=\Delta g^\PZ_1+\tan^2\theta_\PW\Delta\,\kappa_\gamma$}
\end{picture} }
\caption{Influence of the anomalous triple gauge-boson coupling
  $\lambda_\ga$ and of
  non-universal corrections in the $\PW^+$-production-angle
  distribution at $\sqrt{s}=200\GeV$ 
  for the process $\Pep\Pem\to\Pu\bar\Pd\mu^-\bar\nu_\mu$
  (taken from \citere{Denner:2001bd})}
\label{fig:atgc}
\efi
The figure
demonstrates the necessary theoretical improvements. 
At energies beyond $500\GeV$ effects beyond ${\cal O}(\alpha)$,
such as the above-mentioned Sudakov logarithms at higher orders, 
become important and have to be included in predictions at
per-cent accuracy.

At LEP2, the W-boson mass is determined by the reconstruction of the
W~bosons from their decay products with a final accuracy of about $30\MeV$.
In \citere{Jadach:2001cz} the theoretical uncertainty is estimated to be
of the order of $\sim 5\MeV$, which is comparable to the estimated
\cite{talkKM} accuracy of $\sim 10\MeV$ at a future LC.
Theoretical improvements are, thus, desirable.

The main sensitivity of all observables to anomalous couplings in the
triple-gauge-boson vertices is provided by the W-pair production angle
distribution. Figure~\ref{fig:atgc}
shows the impact of an anomalous coupling $\lambda_\gamma=\pm 0.03$,
the size of which is of the order of the LEP2 sensitivity, together with the
impact of non-universal corrections on the spectrum. 
The theoretical uncertainty in constraining the parameter
combination $\la=\la_\gamma=\la_\PZ$ was estimated to be about
$0.005$ \cite{Bruneliere:2002df} for the LEP2 analysis. Since a future
LC is more sensitive to anomalous couplings than LEP2 by more than
an order of magnitude, a further reduction of the uncertainties
by missing radiative corrections is necessary.
However, a thorough estimate of the theoretical uncertainty in the
determination of anomalous couplings at higher scattering energies
($\sqrt{s}\gsim500\GeV$),
where the experimental sensitivity to non-standard couplings increases,
is not yet available.

The above discussion illustrates the necessity of a full one-loop
calculation for the $\Pep\Pem\to 4f$ process
and of further improvements by leading higher-order corrections.

\subsection{Single-W production}

The single-W production process $\Pep\Pem\to\Pe\nu_\Pe\PW\to \Pe\nu_\Pe+2f$
plays a particularly important role among the $4f$ production processes
at high scattering energies. The process is predominantly initiated by
$\Pe\gamma^*$ collision (see \reffi{fig:singleW})
where the photon is radiated off the electron
(or positron) by the Weizs\"acker--Williams mechanism, i.e.\ with
a very small off-shellness $q_\ga^2$. 
\begin{figure}[b]
\centerline{\unitlength 1pt %\small
\begin{picture}(180,90)
\put(30,-10){
\begin{picture}(180,170)(0,0)
\ArrowLine(-10,80)(30,80)
\ArrowLine(30,80)(80,80)
\Photon(30,80)(50,40){2}{5}
\ArrowLine(50,40)(10,10)
\ArrowLine(80,10)(50,40)
\Photon(50,40)(100,40){2}{6}
\ArrowLine(100,40)(140,50)
\ArrowLine(140,30)(100,40)
\GCirc(50,40){10}{.5}
\GCirc(30,80){10}{.5}
\GCirc(100,40){10}{.5}
\put(-7, 3){${\Pep}$}
\put(-28,78){${\Pem}$}
\put(86, 6){${\bar\nu_\Pe}$}
\put(70,46){${W}$}
\put(86,78){${\Pem}$}
\put(46,60){${\gamma}$}
\end{picture}  }
\end{picture}
}
\caption{Generic diagram for the dominant contributions to
single-W production, $\Pep\Pem\to\Pem\nu_\Pe\PW\to \Pem\nu_\Pe+2f$}
\label{fig:singleW}
\end{figure}
Consequently the cross section rises
logarithmically with the scattering energy and is of the same size
as the W-pair production cross section at about $\sqrt{s}=500\GeV$;
for higher energies single-W dominates over W-pair production.

Theoretically the dominance of photon exchange at low $q_\ga^2$ poses
several complications. Technically, $q_\ga^2\to0$ means that the
electrons (or positrons) are produced in the forward direction and
that the electron mass has to be taken into account in order to
describe the cross section there. Moreover, the mere application
of $s$-dependent leading-logarithmic structure functions does not
describe the leading photon-radiation effects
properly, since ISR and final-state radiation (FSR)
show sizeable interferences for forward scattering. Thus, 
the improvement of lowest-order calculations by leading radiation 
effects is more complicated than for $s$-channel-like processes.
Finally, the running of the electromagnetic coupling 
$\alpha(q_\ga^2)$ has to be evaluated in the region of small
momentum transfer ($q_\ga^2<0$) where the fit \cite{Jegerlehner:2001ca}
of this quantity to the hadronic vacuum polarisation should be used. 

The Monte Carlo generator {\sc KoralW} \cite{Jadach:1998gi}
has recently been updated to include the ISR-FSR interference effects
as well as the proper running of $\alpha(q_\ga^2)$.
Therefore, this program now has reached a level of accuracy
similar to the other state-of-the-art programs for single-W production:
{\sc Grc4f} \cite{Fujimoto:1996qg},
{\sc Nextcalibur} \cite{Berends:2000gj},
{\sc Swap} \cite{Montagna:2000ev},
{\sc Wphact} \cite{Accomando:1996es}, and
{\sc Wto} \cite{Passarino:1996rc}.
More detailed descriptions of these codes can be found in
\citere{Grunewald:2000ju}. It should be kept in mind that none
of these calculations includes non-universal electroweak
corrections, leading to a theoretical uncertainty of about
$\sim5\%$ in cross-section predictions. Although the final
solution for a high-energy LC certainly requires a full
${\cal O}(\alpha)$ calculation of the $4f$-production process,
a first step of improvement could be done by a careful
expansion about the propagator poles of the photon and W~boson.
The electroweak ${\cal O}(\alpha)$ corrections to the process
$\Pe\ga\to\nu_\Pe\PW$, which are known \cite{Denner:1992vg},
represent a basic building block in this calculation.
\looseness -1

\subsection{Technical progress on radiative corrections to
multi-particle production processes}

One-loop integrals become more and more cumbersome if the number $N$ of
external legs in diagrams increases. For $N>4$,
however, not all external momenta are linearly independent because of the
four-dimensionality of space-time.
As known for a long time \cite{Me65}, this fact opens the possibility 
to relate integrals with $N>4$ to integrals with $N\le4$.
In recent years, various techniques for actual evaluations of
one-loop integrals with $N=5,6$ have been worked out
\cite{Fleischer:1999hq,Denner:2002ii} 
(see also references therein for older methods and results).
The major complication in the treatment of $2\to3$ processes at one loop
concerns the numerical evaluation of tensor 5-point integrals; in particular,
the occurrence of inverse Gram determinants in the usual
Passarino--Veltman reduction to scalar integrals leads
to numerical instabilities at the phase-space boundary. A possible
solution to this problem was worked out in \citere{Denner:2002ii}
where the known direct reduction 
\cite{Me65} of scalar 5-point to 4-point integrals
was generalized to tensor integrals, thereby avoiding the occurrence
of leading Gram determinants completely.

In the evaluation of real corrections, such as bremsstrahlung,
a proper and numerically stable separation of infrared (soft and collinear)
divergences represents one of the main problems.
In the phase-space slicing approach 
(see \citere{Harris:2001sx} and references therein)
already mentioned above,
the singular regions are excluded from the ``regular'' 
phase-space integration by small cuts on energies, angles, or
invariant masses. Using factorization properties, the integration over
the singular regions can be done in the limit of infinitesimally
small cut parameters. 
The necessary fine-tuning of cut parameters is avoided in so-called
subtraction methods 
(see \citeres{Catani:1997vz,Dittmaier:2000mb,Phaf:2001gc} 
and references therein),
where a specially tuned auxiliary function is subtracted from the 
singular integrand in such a way that the resulting integral is regular.
The auxiliary function has to be chosen simple enough, so that the
singular regions can be integrated over analytically. 
In \citere{Catani:1997vz} the so-called ``dipole subtraction approach''
has been worked out for massless QCD. The technique admits a convenient
construction of such auxiliary functions for arbitrary one-parton
emission processes, without the need of any further complicated
analytical integrations. The dipole subtraction formalism
was subsequently worked out for photon emission off massive
fermions in \citere{Dittmaier:2000mb} and for QCD with massive
quarks in \citere{Phaf:2001gc}.

\subsection{Results on $2\to3$ processes at one-loop order --
$\Pep\Pem\to\nu\bar\nu\PH,\Pt\bar\Pt\PH$}

Recently some one-loop calculations of electroweak radiative corrections
have been presented for $2\to3$ processes that are interesting 
at a future LC: 
$\Pep\Pem\to\nu\bar\nu\PH$ \cite{Belanger:2002ik,Denner:2003yg} and
$\Pep\Pem\to\Pt\bar\Pt\PH$ \cite{You:2003zq,Belanger:2003nm,eetth}.
The results of \citeres{Belanger:2002ik,Belanger:2003nm}
were obtained with the
{\sc Grace-Loop} \cite{Fujimoto:zx} system (see below).
In \citeres{Denner:2003yg,You:2003zq,eetth} the 
technique \cite{Denner:2002ii} for treating tensor 5-point integrals
was employed. 
While \citeres{Belanger:2002ik,You:2003zq,Belanger:2003nm} make
use of the slicing approach for treating soft-photon emission,
the results of \citeres{Denner:2003yg,eetth} have been obtained by
dipole subtraction and checked by phase-space slicing
for soft and collinear bremsstrahlung.

In $\Pep\Pem$ annihilation there are two main production mechanisms
for the SM Higgs boson. In the Higgs-strahlung process,
$\Pep\Pem\to\PZ\PH$, a virtual $\PZ$ boson decays into a $\PZ$ boson
and a Higgs boson. The corresponding cross section rises sharply at
threshold ($\sqrt{s}\gsim\MZ+\MH$)
to a maximum at a few tens of GeV above $\MZ+\MH$ and then
falls off as $s^{-1}$, where $\sqrt{s}$ is the CM
energy of the $\Pep\Pem$ system. In the W-boson-fusion process,
$\Pep\Pem\to\Pne\Pnebar\PH$, the incoming $\Pep$ and $\Pem$ each emit
a virtual W~boson which fuse into a Higgs boson. The cross section of
the W-boson-fusion process grows as $\ln s$ and thus is the dominant
production mechanism for $\sqrt{s}\gg\MH$. 
At the one-loop level, first
the contributions of fermion and sfermion loops in the Minimal
Supersymmetric Standard Model (MSSM)
have been evaluated in \citere{Eberl:2002xd}.
A complete calculation of the $\Oa$ electroweak corrections to $\eennh$
in the SM has subsequently been performed in 
\citeres{Belanger:2002ik,Denner:2003yg}.% 
\footnote{Analytical results for the one-loop corrections have been
obtained in \citere{Jegerlehner:2002es} as {\sc MAPLE} output, but a
numerical evaluation of these results is not yet available.} 
Some results of \citeres{Belanger:2002ik,Denner:2003yg} are 
compared in \refta{tab:eennh}.
\begin{table}
\begin{center}
\caption{Comparison of lowest-order cross sections for $\eennh$,
($\si_{\mathrm{tree}}$), of one-loop-corrected cross section ($\si$),
and of relative corrections ($\delta=\si/\si_{\mathrm{tree}}-1$)
between \citeres{Belanger:2002ik,Denner:2003yg} at $\sqrt{s}=500\GeV$
(input parameters of \citere{Belanger:2002ik})}
\vspace*{.5em}
\begin{tabular}{|c|ccc|c|}
\hline
$\MH[\mathrm{GeV}]$ & $\si_{\mathrm{tree}}[\mathrm{fb}]$ &
$\si[\mathrm{fb}]$ & $\delta[\%]$ & Ref. \\
\hline
150 & 61.074(7)  & 60.99(7)  & $-0.2$ & \cite{Belanger:2002ik} \\
    & 61.076(5)  & 60.80(2)  & $-0.44(3)$ & \cite{Denner:2003yg} \\
\hline
200 & 37.294(4)  & 37.16(4)  & $-0.4$ & \cite{Belanger:2002ik} \\
    & 37.293(3)  & 37.09(2)  & $-0.56(4)$ & \cite{Denner:2003yg} \\
\hline
250 & 21.135(2)  & 20.63(2)  & $-2.5$ & \cite{Belanger:2002ik} \\
    & 21.134(1)  & 20.60(1)  & $-2.53(3)$ & \cite{Denner:2003yg} \\
\hline
300 & 10.758(1)  & 10.30(1)  & $-4.2$ & \cite{Belanger:2002ik} \\
    & 10.7552(7) & 10.282(4) & $-4.40(3)$ & \cite{Denner:2003yg} \\
%\hline
%350 & 4.6079(5)  & 4.184(4)  & $-9.1$ & \cite{Belanger:2002ik} \\
%    & 4.6077(2)  & 4.181(1)  & $-9.27(3)$ & \cite{Denner:2003yg} \\
\hline
\end{tabular}
\label{tab:eennh}
\end{center}
\end{table}
The agreement of the correction is within 0.2\% or better w.r.t.\
the lowest-order cross sections.

The Yukawa coupling of the top quark could be measured at a future
LC with high energy and luminosity at the level of $\sim5\%$ 
\cite{Aguilar-Saavedra:2001rg} by analyzing the process $\eetth$.
A thorough prediction for this process, thus, has to control QCD
and electroweak corrections. 
Table~\ref{tab:eetth} summarizes some results on the electroweak
$\Oa$ corrections of \citeres{Belanger:2003nm,eetth}.
\begin{table}
\bce
\caption{Comparison of lowest-order cross sections for $\eetth$,
($\si_{\mathrm{tree}}$), of one-loop-corrected cross section ($\si$),
and of relative corrections ($\delta=\si/\si_{\mathrm{tree}}-1$)
between \citeres{Belanger:2003nm,eetth} for $\MH=120\GeV$
(input parameters of \citere{Belanger:2003nm},
results taken from \citere{eetth})}
\label{tab:eetth}
\vspace*{.5em}
\begin{tabular}{|c|ccc|c|}
\hline
$\sqrt{s}[\mathrm{GeV}]$ & $\si_{\mathrm{tree}}[\mathrm{fb}]$ &
$\si[\mathrm{fb}]$ & $\delta[\%]$ & Ref. \\
\hline
 600 & 1.7293(3) & 1.738(2)  & 0.5     & \cite{Belanger:2003nm} \\
     & 1.7292(2) & 1.7368(6) & 0.44(3) & \cite{eetth} \\
\hline
 800 & 2.2724(5) & 2.362(4)  & 3.9     & \cite{Belanger:2003nm}\\
     & 2.2723(3) & 2.3599(6) & 3.86(2) & \cite{eetth} \\
\hline
1000 & 1.9273(5) & 2.027(4)  & 5.2     & \cite{Belanger:2003nm}\\
     & 1.9271(3) & 2.0252(5) & 5.09(2) & \cite{eetth} \\
\hline
\end{tabular}
\ece
\end{table}
The agreement within $\sim 0.1\%$ also holds for other energies
and Higgs-boson masses.
The results of the previous calculation \cite{You:2003zq}
roughly agree with the ones of \citeres{Belanger:2003nm,eetth} at
intermediate values of $\sqrt{s}$ and $\MH$, but are at variance
at high energies (TeV range) and close to threshold (large $\MH$).

\section{EVENT GENERATORS FOR MULTI-PARTICLE FINAL STATES}

\subsection{Multi-purpose generators at parton level}

The large variety of different final states for multi-particle
production renders multi-purpose Monte Carlo event generators rather 
important, i.e.\ generators that deliver
an event generator for a user-specified (as much as possible)
general final state based on full lowest-order amplitudes.
As results, these tools yield lowest-order predictions for
observables, or more generally Monte Carlos samples of events,
that are improved by universal radiative corrections, such as
initial-state radiation at the leading-logarithmic level or 
beamstrahlung effects.
Most of the multi-purpose generators are also interfaced to parton-shower
and hadronization programs.
The generality renders these programs, however, rather complex devices
and, at present, they are far from representing tools for high-precision
physics, because non-universal radiative corrections are not taken 
into account in predictions.

The following multi-purpose generators for multi-parton production, 
including program packages for the matrix-element evaluation, are available:
\begin{itemize}
\item
{\sc Amegic} \cite{Krauss:2001iv}: 
Helicity amplitudes are automatically generated by the program for
the SM, the MSSM, and some new-physics models. Various interfaces
(ISR, PDFs, beam spectra, {\sc Isajet}, etc.) are supported.
The phase-space generation was successfully tested for up to
six particles in the final state.
\item
{\sc Grace} \cite{Ishikawa:1993qr}: The amplitudes are delivered
by a built-in package, which can also handle SUSY processes. The
phase-space integration is done by {\sc BASES} \cite{Kawabata:1995th}.
Tree-level calculations have been performed for up to 
(selected) six-fermion final states. The extension of the system
to include one-loop corrections, the {\sc Grace-Loop} \cite{Fujimoto:zx}
program, is under construction.
\item
{\sc Madevent} \cite{Maltoni:2002qb} + {\sc Madgraph} \cite{Stelzer:1994ta}:
The {\sc Madgraph} algorithm can generate tree-level matrix elements
for any SM process (fully supporting particle masses), but a practical 
limitation is 9,999 diagrams.
In addition, {\sc Madgraph} creates {\sc Madevent}, an event generator for 
the requested process. 
\item
{\sc Phegas} \cite{Papadopoulos:2000tt} + {\sc Helac} \cite{Kanaki:2000ey}:
The {\sc Helac} program delivers amplitudes for all SM processes 
(including all masses). The phase-space integration done by {\sc Phegas}
has been tested for selected final states with up to seven particles.
\item
{\sc Whizard} \cite{Kilian:2001qz} 
+ {\sc Comphep} \cite{Pukhov:1999gg} / {\sc Madgraph} \cite{Stelzer:1994ta}
/ {\sc O'mega} \cite{Moretti:2001zz}:
Matrix elements are generated by an automatic interface to (older versions of)
{\sc Comphep}, {\sc Madgraph}, and (the up-to-date version of) {\sc O'mega}.
Phase-space generation has been tested for most $2\to6$ and some
$2\to8$ processes; unweighed events are supported, and a large variety
of interfaces (ISR, beamstrahlung, {\sc Pythia}, PDFs, etc.) exists.
The inclusion of MSSM amplitudes ({\sc O'mega}) and improved
phase-space generation ($2\to6$) are work in progress.
\end{itemize}
All but the {\sc Grace} program make use of the multi-channel approach
for the phase-space integration. More details can be found in the
original references.

Tuned comparisons of different generators, both at parton and 
detector level, are extremely important, but become more and more
laborious owing to the large variety of multi-particle final states.
Some progress to a facilitation and automization of comparisons 
are made by MC-tester project \cite{Golonka:2002rz} and 
Java interfaces \cite{Ronan:2003wq}.

\subsection{Event generators and results for $\Pep\Pem\to6f$}

Particular progress was reached in recent years in the description
of six-fermion production processes. Apart from the multi-purpose generators
listed in the previous section, also dedicated Monte Carlo programs
and generators have been developed for this class of processes:
\begin{itemize}
\item
{\sc Sixfap} \cite{Montagna:1997dc}: Matrix elements are provided
for all $6f$ final states (with finite fermion masses), including
all electroweak diagrams. The generalization to QCD diagrams and the
extension of the phase-space integration for all final states is in
progress.
\item
{\sc eett6f} \cite{Kolodziej:2001xe}: Only processes relevant for
$\Pt\bar\Pt$ production are supported (a new version includes
$\Pe^\pm$ in the
final state and QCD diagrams); finite fermion masses are possible. 
\item
{\sc Lusifer} \cite{Dittmaier:2002ap}: All $6f$ final states are
possible, including QCD diagrams with up to four quarks; 
representative results for all these final states have been presented.
External fermions are massless.
An unweighting algorithm and an interface to {\sc Pythia} are available.
\end{itemize}
\begin{table*}
\caption{Comparison of lowest-order predictions for some processes
$\Pep\Pem\to\Pt\bar\Pt\to6\,$fermions at $\sqrt{s}=500\GeV$ in
approximation of massless fermions
(input parameters and cuts of \citere{Dittmaier:2002ap})}
\vspace*{.5em}
\centerline{
\begin{tabular}{|l|cccccc|}
\hline
$\si_{\mathrm{full}}[\fb]$ &
{\sc Amegic++} &
{\sc eett6f} &
{\sc Lusifer}&
{\sc Phegas} &
{\sc Sixfap} &
{\sc Whizard} \\
\hline
$\nu_\Pe\Pe^+\Pe^-\bar\nu_\Pe\Pb\bar\Pb$ &
        5.879(8) & 5.862(6) & 5.853(7) & 5.866(9) & 5.854(3) & 5.875(3) \\
$\nu_\Pe\Pe^+\mu^-\bar\nu_\mu\Pb\bar\Pb$ &
        5.827(4) & 5.815(5) & 5.819(5) & 5.822(7) & 5.815(2) & 5.827(3) \\
$\nu_\mu\mu^+\mu^-\bar\nu_\mu\Pb\bar\Pb$ &
        5.809(5) & 5.807(3) & 5.809(5) & 5.809(5) & 5.804(2) & 5.810(3) \\
$\nu_\mu\mu^+\tau^-\bar\nu_\tau\Pb\bar\Pb$ &
        5.800(3) & 5.797(5) & 5.800(4) & 5.798(4) & 5.798(2) & 5.796(3) \\
$\nu_\mu\mu^+\Pd\bar\Pu\Pb\bar\Pb$ &
        17.209(9) & 17.213(23) & 17.171(24) & 17.204(18) &  & \\[-.3em]
without QCD: &
        17.097(8) & 17.106(15) & 17.095(11) & 17.107(18) & 17.096(4) & 17.103(8) \\
\hline
\end{tabular} }
\label{tab:eett6f}
\end{table*}
Table~\ref{tab:eett6f} summarizes a brief comparison of results
for some processes $\Pep\Pem\to 6f$ relevant for $\Pt\bar\Pt$ production
for massless external fermions.
The results reveal good agreement between the various programs, where
minor differences are presumably due to the different treatments of the 
bottom-quark Yukawa coupling, which is neglected in some cases.

A tuned comparison of results obtained with {\sc Lusifer} and
{\sc Whizard} for a large survey of $6f$ final states has been
presented in \citere{Dittmaier:2002ap}.

\subsection{``True'' Monte Carlo event generators}

The event generators described above work at parton level 
(partially improved by parton showers), i.e.\ the final-state particles 
cannot be directly identified with particles in a detector. 
For detector simulations, these parton-level generators have to be
interfaced with parton shower and hadronization programs.
To facilitate this interface, the ``Les Houches accord'' 
\cite{Boos:2001cv} has been designed, a set
of {\sc Fortran} common blocks for the transfer of event configurations 
from parton level generators to showering and hadronization event generators.
Alternatively
so-called ``true'' event generators could be used,
which fully include hadronization.
The following well-known ``true'' MC generators represent general-purpose tools
for investigating not only $\Pep\Pem$ collisions, but also lepton--hadron
and hadron--hadron scattering:
{\sc Herwig} \cite{Moretti:2002rj},
{\sc Isajet} \cite{Baer:1999sp}, and
{\sc Pythia} \cite{Sjostrand:2001yu}.
The programs are supported and extended continuously; recent upgrades
and new features relevant for LC physics are:
\begin{itemize}
\item
implementation of all $2\to2$ scattering processes of the MSSM in lowest order;
\item
associated Higgs production ($Q\bar{Q}^{(\prime)}H$) in the SM and MSSM 
in lowest order (the case of charged Higgs bosons is not yet 
available in {\sc Isajet} and {\sc Pythia});
\item
$R$-parity-violation SUSY in {\sc Herwig};
\item
MSSM with complex parameters (cMSSM) in {\sc Pythia};
\item
inclusion of spin correlations and matrix elements for 3- and 4-body decays
in {\sc Herwig};
\item
introduction of real corrections based on matrix elements
for several $\Pep\Pem$ processes in {\sc Herwig} and {\sc Pythia}
(see e.g.\ \citere{Andre:1997zf}).
\end{itemize}
Among other work in progress, the implementation of NLO QCD corrections
in ``true'' generators is one of the most pressing issues.
In particular, the matching of parton showers with 
matrix-element calculations at NLO has to be performed carefully;
first results look very promising \cite{Frixione:2002ik}.

Finally, it should be mentioned that the present {\sc Fortran}
versions of {\sc Herwig} and {\sc Pythia} will be replaced by
{\sc C++} programs in the future \cite{Bertini:2000uh}.

\section{RADIATIVE CORRECTIONS IN SUPERSYMMETRIC THEORIES}

In order to avoid too much overlap with the reports of the 
Higgs and SUSY groups, this section is mainly restricted to 
the topics that have 
been presented in the Loopverein working group.
More details and references on the subject can be found in
\citere{theseprocs}.

\subsection{SUSY corrections to precision observables}

\begin{figure}
{ \setlength{\unitlength}{1cm}
\begin{picture}(8,8.0)
\put(-0.4,-0.9){\includegraphics{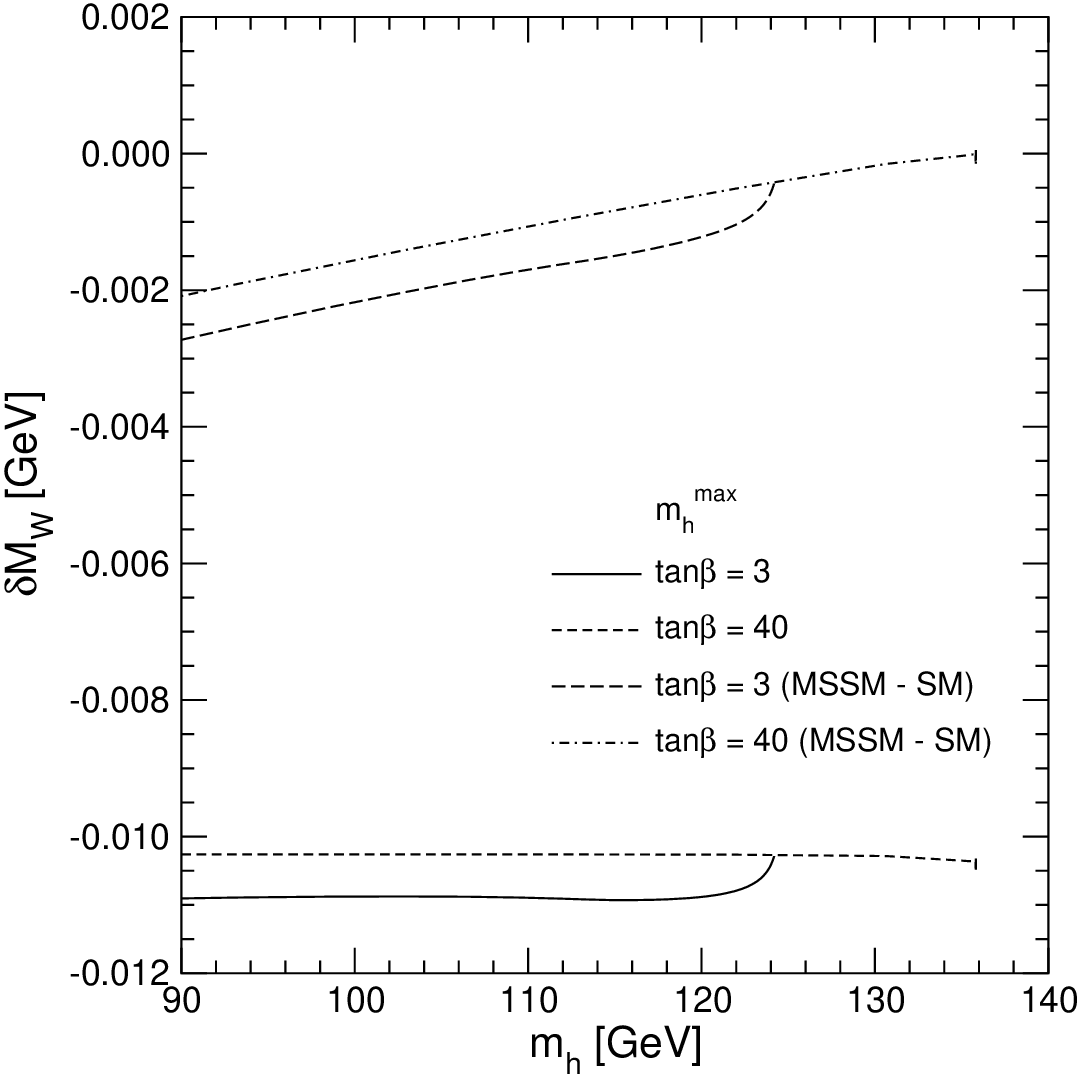}}
\end{picture} }
\caption{Effect of the SUSY
${\cal O}(\alpha_{\Pt}^2)$ corrections on the prediction of $\MW$ in the MSSM
(taken from \citere{Heinemeyer:2002jq})}
\label{fig:mwpredmssm}
\end{figure}

The confrontation of high-precision data with theoretical predictions
is, of course, also very interesting in extensions of the SM. 
The one-loop corrections of the MSSM to muon decay and to the 
pseudo-observables of the Z~resonance have been known for many years, but
not many corrections beyond one loop exist. 
Recently the known universal corrections of
${\cal O}(\alpha\alpha_{\mathrm{s}})$ \cite{Djouadi:1998sq}
to the $\rho$-parameter have been supplemented by the terms of
${\cal O}(\alpha_{\Pt}^2)$,
${\cal O}(\alpha_{\Pt}\alpha_{\Pb})$
and ${\cal O}(\alpha_{\Pb}^2)$ in \citere{Heinemeyer:2002jq},
where $\alpha_{\Pt/\Pb}=h_{\Pt/\Pb}^2/(4\pi)$ 
with $h_{\Pt/\Pb}$ denoting the top/bottom Yukawa coupling.
Figure~\ref{fig:mwpredmssm} illustrates the effect of the
${\cal O}(\alpha_{\Pt}^2)$ corrections on the prediction of the W-boson mass
in the MSSM.
The genuine MSSM ${\cal O}(\alpha_{\Pt}^2)$ 
effects modify $\MW$ at the level of $2{-}3\MeV$.

\subsection{Mass spectra in the MSSM}

In theories with unbroken supersymmetry the fermions and bosons within the 
same multiplet have a common mass. In realistic theories, such as the MSSM, 
SUSY is broken, and this statement is not valid anymore.
However, the masses of fermions or bosons within multiplets are 
not all independent, i.e.\ there are
non-trivial relations among mass parameters.
Since the mass spectra of SUSY theories bear a lot of information on the
mechanism of SUSY breaking, precision analyses of these spectra can
serve as a window to grand unification.

SUSY demands (at least) two Higgs doublets to give the up- and down-type
fermions masses. Thus, the MSSM predicts the existence of two charged
($\PH^\pm$), two neutral scalar ($\Ph^0,\PH^0$), and one neutral pseudo-scalar
($\PA^0$) Higgs bosons. In lowest order, the Higgs masses $M_{\PH^\pm}$,
$M_{\Ph}$, and $M_{\PH}$ can be calculated as functions of the
$\PA^0$-boson mass $M_\PA$, the ratio of Higgs-field vacuum expectation
values, $\tan\beta=v_2/v_1$, and the gauge-boson masses; in particular,
the mass $M_{\Ph}$ of the lightest Higgs boson is constrained to be 
smaller than $\MZ$ at tree level. 
Beyond lowest order, also the remaining MSSM parameters are 
involved in the mass relations, and $M_{\Ph}$ can reach values up to
about $135\GeV$.
The status of precision calculations of the neutral Higgs-boson
masses has been recently reviewed in \citere{Degrassi:2002fi}.
All available corrections%
\footnote{It should be mentioned that the full effective potential has
been presented in \citere{Martin:2002iu} at the two-loop level. However,
the precise relation of these results with the parameters defined in 
the on-shell renormalization scheme used in MSSM parameter analyses has not
been worked out so far.}
are implemented in the program
{\sc FeynHiggs} \cite{Heinemeyer:1998yj}.
The predictions are based on full one-loop calculations and on the
leading effects in two-loop order, i.e.\ the corrections of the
order ${\cal O}(\alpha_{\mathrm{s}}\alpha_{\Pt})$,
${\cal O}(\alpha_{\Pt}^2)$ 
and ${\cal O}(\alpha_{\mathrm{s}}\alpha_{\Pb})$
(see \citere{Degrassi:2002fi} and references therein).
Recently the corrections of ${\cal O}(\alpha_{\Pt}\alpha_{\Pb})$ 
and ${\cal O}(\alpha_{\Pb}^2)$ have become available \cite{Dedes:2003km}.
The current theoretical uncertainty in the Higgs-mass predictions
is about $3\GeV$ \cite{Degrassi:2002fi}, 
but a further reduction to $\lsim 0.5\GeV$
should be reached to match the accuracy needed for a LC.
In this context a proper definition of $\tan\beta$ in higher perturbative
orders is crucial, since different renormalization schemes 
(see, e.g., \citere{Freitas:2002um})
for $\tan\beta$ lead to rather different relations between $\tan\beta$
and physical observables such as the Higgs-boson masses.

In the sector of charginos and neutralinos of the MSSM only three out
of the six mass parameters are independent. 
In \citeres{Eberl:2001eu,Fritzsche:2002bi} the three masses
$m_{\tilde\chi^0_{2,3,4}}$ of the heavier neutralinos have been 
expressed in terms of the mass $m_{\tilde\chi^0_{1}}$ of the lightest 
neutralino and of the masses $m_{\tilde\chi^+_{1,2}}$ of the two
charginos, including the complete one-loop corrections, which depend
also on the other MSSM parameters. The corrections modify the
calculated masses by up to several GeV. 
The on-shell renormalization in the chargino-neutralino sector 
is described in \citeres{Eberl:2001eu,Fritzsche:2002bi,Guasch:2001kz}
in detail.

The relations among sfermion masses, together with the corresponding
on-shell renormalization, are worked out in 
\citeres{Guasch:2001kz,Hollik:2003jj}.
For each generation, one of the four squark masses and one of the three
slepton masses can be calculated from the other sfermion masses 
(and the other MSSM parameters entering at one loop). The one-loop
corrections can amount to about 5\%.

\subsection{Higgs-boson and SUSY-particle decays in the MSSM}

Analyses of particle decays are of great importance for the reconstruction 
of coupling structures and, thus, of interaction Lagrangians. 
The rich particle content of SUSY theories leads to a large variety
of decay cascades, which depend in detail on the chosen scenario.
A discussion of phenomenological implications and of radiative
corrections to SUSY-particle decays can be found in \citere{theseprocs}
and \citere{Majerotto:2002iu}, respectively.

The decay widths of Higgs bosons in the MSSM received much attention
in recent years, so that the predictions are well elaborate.
Precise predictions can be obtained with the programs
{\sc FeynHiggsDecay} (based on \citere{Heinemeyer:2000fa}) and
{\sc Hdecay} \cite{Djouadi:1997yw}. 
Recently the electroweak ${\cal O}(\alpha)$ corrections to
the decay of the CP-odd $\PA^0$ boson into sfermion pairs
have been calculated in \citere{Weber:2002su}.
For the Higgs decay $\phi\to\Pb\bar\Pb$ ($\phi=\Ph^0,\PH^0,\PA^0$), 
which is of particular
importance for light Higgs bosons, the resummation of the leading
SUSY-QCD effects and the related theoretical uncertainty 
have been discussed in \citere{Guasch:2003cv} 
(for previous work on $\phi\to\Pb\bar\Pb$ see references therein).

Apart from considering integrated decay rates, it is interesting to
inspect distributions of decay products, which is important for
the determination of the spin and parity of the decaying particle.
This task requires the development of appropriate Monte Carlo
tools. For the $\phi\to\tau^+\tau^-$ decay, for instance, the
{\sc Tauola} program was extended to include $\tau$-spin correlations in
\citere{Was:2002gv}.

\subsection{SUSY-particle production}

The direct production of SUSY particles, if they exist, is among
the most interesting issues at future colliders. In order to 
determine the properties (mass, spin, decay widths, couplings)
of these new particles, precise measurements and predictions of the
corresponding cross sections at the same level of accuracy 
are necessary, i.e.\ radiative
corrections have to be taken into account.

The electroweak radiative corrections to the production of
sfermions, $\Pep\Pem\to\tilde f\bar{\tilde f}$,
and charginos, $\Pep\Pem\to\tilde\chi^+\tilde\chi^-$,
were worked out in \citeres{Freitas:2002ge}
and \cite{Blank:2000uc},
respectively. Since in these calculations the sfermions and charginos 
are assumed to be stable, the results are relevant for energies a few 
decay widths above the production threshold. For a threshold scan the 
decay of the sfermions as well as non-resonant coherent background effects
have to be included; 
in \citere{Freitas:2001zh} off-shell effects in sfermion-pair 
production have been investigated at tree level (improved by
universal Coulomb-like corrections).
Theoretically the whole issue is very similar
to a description of the process $\Pep\Pem\to\PW\PW\to 4f$ which
is discussed above.

The SUSY multiplet structure, in particular, predicts that the
strengths of the gauge-boson--fermion and
gauge-boson--sfermion interactions, which are equal owing to the gauge
principle, coincide with the gaugino--sfermion--fermion Yukawa coupling.
In order to test this relation in SUSY QCD the processes 
$\Pep\Pem\to q\bar{q}\Pg, \tilde q\bar{\tilde q}\Pg,
\tilde q\bar{q}\tilde\Pg, q\bar{\tilde q}\tilde\Pg$
should be studied. 
In \citere{Brandenburg:2002ff} the QCD and SUSY-QCD corrections
to these processes were calculated.

More details on radiative corrections to SUSY particle production
and decays
can be found in \citere{Majerotto:2002iu} (and references therein).

\subsection{Renormalization of the MSSM beyond one loop}

Beyond one loop the calculation of radiative corrections within
SUSY theories is highly non-trivial, because there is no
regularization scheme that respects gauge invariance and
supersymmetry at the same time. For instance, conventional
dimensional regularization breaks supersymmetry, while dimensional
reduction, which respects supersymmetry, is known to be not
fully consistent. In this situation, a mathematically convincing 
way to perform renormalization is provided by {\it algebraic
renormalization}. In this framework the symmetry identities and
a proof of renormalizability
for the MSSM have been established in \citere{Hollik:2002mv}.
These results can serve as a basis for the construction of all
symmetry-restoring counterterms in the MSSM.

\section{OTHER DEVELOPMENTS}

\subsection{Automization of loop calculations and maintenance of 
computer codes}

Once the necessary techniques and theoretical subtleties of a 
perturbative calculation are settled, to carry out the actual
calculation is an algorithmic matter. Thus, an automization of
such calculations is highly desirable, in order to facilitate
related calculations. Various program packages
have been presented in the literature for automatized tree-level,
one-loop, and multi-loop calculations. A comprehensive
overview was, for instance, given in \citere{Harlander:1998dq};
in the following we have to restrict ourselves to a selection of
topics, where the emphasis is put on recent developments.

The generation of Feynman
graphs and amplitudes is a combinatorial problem
that can be attacked with computer algebra.
The program packages {\sc FeynArts} \cite{Kublbeck:1990xc}
(which has been extended \cite{Hahn:2001rv} for the MSSM),
{\sc Qgraf} \cite{Nogueira:1991ex}, {\sc Diana} \cite{Tentyukov:1999is}
(based on {\sc Qgraf}) are specifically designed for this task; also the 
{\sc Grace-Loop} \cite{Fujimoto:zx} system automatically generates
Feynman diagrams and loop amplitudes. 
Moreover, the task of calculating
virtual one-loop and the corresponding real-emission corrections
to $2\to 2$ scattering reactions is by now well understood.
Such calculations are widely automated in the packages
{\sc FormCalc}, combined with {\sc LoopTools} \cite{Hahn:1998yk},
and {\sc Grace-Loop} \cite{Fujimoto:zx}.

As an illustrating example, \refta{tab:eett} provides some results
on the differential cross section for $\Pep\Pem\to\Pt\bar\Pt$
in lowest order as well as including electroweak ${\cal O}(\alpha)$
corrections.
\def\sig{\left(\frac{\displaystyle{\mathrm{d}\sigma}}{\displaystyle{\mathrm{d}\cos \, \theta}}\right)}
\begin{table*}
\caption{
Differential cross sections for $\Pep\Pem\to\Pt\bar\Pt$
for selected scattering angles at $\sqrt{s}=500\GeV$;
input parameters are defined in 
\citere{Fleischer:2002rn},
%\citere{Fleischer:2003kk,LC-TH-2003-035,LC-TH-2003-036},
the soft-photon cut parameter $\omega/\sqrt{s}$ is set to $10^{-5}$.}
\label{tab:eett}
\vspace*{.5em}
\centerline{$
\begin{array}{|r|l|l|l|l|}
\hline 
\vrule height 3ex depth 0ex width 0ex
\cos\theta & \mbox{program} & \sig_{\mathrm{Born}}[\pba]  & 
\sig_{\mathrm{Born+virt+soft}}[\pba] & \sig_{\mathrm{Born+virt+real}}[\pba] \\ 
\hline  
-0.9 & \mbox{\sc FA + FC}    & 0.108839194076039 & -0.00205485893415 &\\
     & \mbox{\sc Grace-Loop} & 0.108839194076    & -0.002054859      & 0.13206(12)\\
     & \mbox{\sc Sanc}       & 0.10883919407522  & -0.00205485893360 &\\
     & \mbox{\sc Topfit}     & 0.108839194076039 & -0.00205485893466 & 0.13229 \\
\hline 
0.0 & \mbox{\sc FA+FC}      & 0.225470464033559  & -0.04321416793299 &\\ 
    & \mbox{\sc Grace-Loop} & 0.225470464033     & -0.043214168      & 0.23513(14)\\
    & \mbox{\sc Sanc}       & 0.22547046403258   & -0.04321416793300 &\\ 
    & \mbox{\sc Topfit}     & 0.225470464033559  & -0.04321416793192 & 0.23476\\
\hline 
+0.9 & \mbox{\sc FA+FC}      & 0.491143715767761 & -0.16747885864057 &\\
     & \mbox{\sc Grace-Loop} & 0.491143715767    & -0.16747886       & 0.47709(21)\\
     & \mbox{\sc Sanc}       & 0.49114371576694  & -0.16747885864510 &\\
     & \mbox{\sc Topfit}     & 0.491143715767761 & -0.16747885863793 & 0.47768
\\ \hline
\end{array}
$}
\end{table*}
The program {\sc FA}+{\sc FC} was obtained from the output of the
{\sc FeynArts} and {\sc FormCalc} packages and makes use of the
{\sc LoopTools} library for the numerical evaluation.
The {\sc Topfit} program \cite{Fleischer:2002rn,Fleischer:2003kk}
was developed from an algebraic reduction of Feynman graphs 
(delivered from {\sc Diana}) within {\sc Form}; for the
numerics {\sc LoopTools} is partially employed. 
More detailed comparisons between {\sc FA}+{\sc FC} and {\sc Topfit}, 
including also other fermion flavours, 
can be found in \citeres{Fleischer:2002rn,Hahn:2003ab}.
The {\sc Grace-Loop} result is completely independent of the two
others. The agreement between these results reflects the enormous
progress achieved in recent years in the automization of
one-loop calculations.

The {\sc Grace-Loop} system has recently been used in the
calculation of the electroweak corrections to the $2\to3$ processes
$\Pep\Pem\to\nu\bar\nu\PH,\Pt\bar\Pt\PH$
\cite{Belanger:2002ik,Belanger:2003nm}, which are discussed above.

Clearly the calculation of radiative corrections is a very laborious task,
leading to rather complex and lengthy computer codes, which should be
carefully documented. 
The {\sc Sanc} project \cite{Andonov:2002jg} 
(former {\sc CalcPHEP} \cite{Andonov:2002rr})
aims at providing theoretical support of this kind for future
accelerator experiments, using the principle of knowledge storing.
This approach is rather different from the strategy of automization
described above, which aims at generating completely new programs.
The {\sc Sanc} program contains another 
independent calculation of the ${\cal O}(\alpha)$
corrections to $\Pep\Pem\to\Pt\bar\Pt$, the results of which are also
included in \refta{tab:eett}.

\subsection{Numerical approaches to loop calculations}

Most of the various techniques of performing loop calculations share
the common feature that the integration over the loop momenta is 
performed analytically. This procedure leads to complications
at one loop if five or more external legs are involved, since
both speed and stability of programs become more and more 
jeopardized. At the two-loop level, already the evaluation of 
self-energy and vertex corrections can lead to extremely complicated
higher transcendental functions that are hard to evaluate numerically.

An idea to avoid these complications is provided by a more or less
purely numerical evaluation of loop corrections. There are two main
difficulties in this approach. Firstly, the appearing ultraviolet and
infrared divergences have to be treated and canceled carefully. 
Secondly, even finite loop integrals require a singularity handling
of the integrand near so-called particle poles, where Feynman's
$\ri\eps$ prescription is used as regularization.

In \citere{Passarino:2001wv} a method for a purely numerical evaluation of
loop integrals is proposed. Each integral is parametrized with 
Feynman parameters and subsequently rewritten with partial
integrations. The final expression consists of a quite simple part 
containing the singular terms and another more complicated looking
part that can be integrated numerically. 
The actual application of the method to a physical process is still
work in progress.

Another idea was proposed in \citere{Soper:2001hu} and applied to
event-shape variables in $\Pep\Pem\to3\,$jets in NLO.
In this approach virtual and real
corrections are added before integration. In their sum, no soft
singularities, or more generally singularities that cancel between
virtual and real corrections, appear from the beginning.
Nevertheless the problem of a stable treatment of particle poles
in loop amplitudes still remains. In \citere{Soper:2001hu} a solution
via contour deformations in complex integration domains is described,
but how this procedure can be generalized is not yet clear.

\section{CONCLUSIONS}

In spite of the complexity of higher-order calculations for 
high-energy elementary particle reactions, there has been 
continuous progress in the development of new techniques and in
making precise predictions for physics at future colliders.
However, to be prepared for a future $\Pep\Pem$ linear collider
with high energy and luminosity, such as TESLA, an enormous
amount of work is still ahead of us. Full two-loop predictions
for $\Pep\Pem\to 2\,$fermion scattering reactions, 
such as the Bhabha process, or full one-loop calculations for
$\Pep\Pem\to 4\,$fermions are more than technical challenges.
At this level of accuracy, field-theoretical issues such as
renormalization, the treatment of unstable particles, etc.,
demand a higher level of understanding. 
Of course, both loop calculations as well as the descriptions
of multi-particle production processes with Monte Carlo
techniques require and will profit from further improving
computing devices.

It is certainly out of question that the list of challenges 
and interesting issues could be continued at will. 
The way to a future LC will also be highly exciting in precision physics.

%\begin{thebibliography}{9}   % Use for  1-9  references

\end{document}